\documentclass[aps,preprint,showpacs, showkeys]{revtex4}%
\usepackage{amsfonts}
\usepackage{amsmath}
\usepackage{amssymb}
\usepackage{graphicx}%
\providecommand{\U}[1]{\protect \rule{.1in}{.1in}}

\begin{document}
\title[ ]{Representation Dependence of Superficial Degree of Divergences in Quantum
Field Theory }
\author{Abouzeid. M. Shalaby\footnote{E-mail:amshalab@mans.edu.eg}}
\affiliation{Physics Department, Faculty of Science, Mansoura University, Egypt}

\begin{abstract}
In this work, we investigate a very important but unstressed result in the
work of Carl M. Bender, Jun-Hua Chen, and Kimball A. Milton (
J.Phys.A39:1657-1668, 2006). In this article, Bender \textit{et.al} have
calculated the vacuum energy of the $i\phi^{3}$ scalar field theory and its
Hermitian equivalent theory up to $g^{4}$ order of calculations. While all the
Feynman diagrams of the $i\phi^{3}$ theory are finite in $0+1$ space-time
dimensions, some of the corresponding Feynman diagrams in the equivalent
Hermitian theory are divergent. In this work, we show that the divergences in
the Hermitian theory originate from superrenormalizable, renormalizable and
non-renormalizable terms in the interaction Hamiltonian even though the
calculations are carried out in the $0+1$ space-time dimensions. Relying on
this interesting result, we raise the question, is the superficial degree of
divergence of a theory is representation dependent? To answer this question,
we introduce and study a class of non-Hermitian quantum field theories
characterized by a field derivative interaction Hamiltonian. We showed that
the class is physically acceptable by finding the corresponding class of
metric operators in a closed form. We realized that the obtained equivalent
Hermitian and the introduced non-Hermitian representations have coupling
constants of different mass dimensions which may be considered as a clue for
the possibility of considering non-Renormalizability of a field theory as a
non-genuine problem. Besides, the metric operator is supposed to disappear
from path integral calculations which means that physical amplitudes can be
fully obtained in the simpler non-Hermitian representation.

\end{abstract}
\maketitle

Two main reasons prevent a theory from playing a role in the description of
matter interactions. In a historical order, the first reason is
non-Hermiticity of Hamiltonian models which for long time prevented any try to
take non-Hermitian theories into account in the search for a suitable
mathematical description that mimic known features of nature. For instance, in
the standard model for particle interactions one had to resort to a
non-Abelean theory of high group structures to obtain the asymptotic freedom
property (QCD). Recently, an idea back to Symanzik has been stressed and it is
now well known that a simple Abelean theory can exhibit asymptotic freedom
\cite{Frieder,symanzic,symanzic1,symanzic2,bendr}. Moreover, also in the
standard model, Hemriticity obliged us to employ a spontaneous symmetry
breaking algorithm using only Hermitian scalar field theory which leaded to
the famous Hierarchy problem. In Ref. \cite{aboebt}, we showed that the
behavior of a non-Hermitian scalar field theory at high energy scales is
secure rather than the Hermitian scalar field theory for which the mass
parameter and all the dimensionfull parameters blow up to unacceptable values
and thus leading to the Hierarchy and the cosmological constant problems
\footnote{In fact, it is a Hierarchy problem too.}. Rather than these
interesting features of a non-Hermitian theory, the techniques used by the
field of pseudo-Hermitian theories can be employed to solve some problems in
Physics. For instance, in Ref.\cite{ghost}, we showed that the algorithm can
be used to cure the ghost states in the Lee-Wick standard model. In fact,
this trend has been initiated by Bender \textit{et.al} in Refs.\cite{ghost1,
ghost2}.

The renormalizability of a theory is the second reason that prevents a theory
from playing a role in describing a physical system. For instance, the
particle physics community celebrated the discovery of the Weinberg-Salam
model for its renormalizability and was ready to replace the
non-Renormalizable Fermi model introduced to describe Weak interactions.
Another famous problem is the unification of the four forces which up till now
is far from reaching a suitable treatment due to the non-Renormalizability of
the theory describing gravitational interactions. Besides, in the early
universe studies, one needs to resort to a theory which is capable of making a
strong first order phase transition to account for matter-antimatter asymmetry
in the universe \cite{early}. A well known theory that can have such feature
is the non-renormalizable $\phi^{6}$ scalar field theory \cite{early2}.

The first reason mentioned above is no longer a  holy belief and one can show
that there exists an infinite number of Hamiltonians which are neither
Hermitian nor $\mathcal{PT}$-symmetric and have real spectra as well. On the
other hand, normalizability have not been stressed in the sense that there are
no known technique by which one can get rid of it. In this work we argue that
a question of the form "is non-renormalizability of a theory is a technical or
a conceptual problem?" is legal. Also, one may ask if it is technical then
what is the calculational algorithm that can be used to get rid of it?
However, before trying to answer this question, we refer to an unstressed but
interesting result in Ref. \cite{bendvs}. In this article, the authors have
obtained the vacuum energy of the $\mathcal{PT}$-symmetric $\left(  i\phi
^{3}\right)  $ scalar field theory which ought to be the same as the vacuum
energy of the corresponding equivalent Hermitian theory. Although the
$\mathcal{PT}$-symmetric $\left(  i\phi^{3}\right)  $ is finite in $0+1$ space
time dimensions, the authors found that the Hermitian theory suffers from the
existence of divergences. This interesting result pushed us to ask the above
question because the results in Ref. \cite{bendvs} show the equivalence
between finite theory and a theory with divergences. \ In fact, Bender
\textit{et.al }showed that the finite non-Hermitian theory described by the
Hamiltonian density of the form;%
\[
H=\frac{\pi^{2}}{2}+\frac{1}{2}\left(  \nabla \phi \right)  ^{2}+\frac{1}%
{2}m^{2}\phi^{2}+ig\phi^{3},
\]
in $0+1$ dimensions of the space-time is equivalent to the divergent Hermitian
Hamiltonian of the form;
\begin{align}
h  &  =\frac{\pi^{2}}{2}+\frac{1}{2}m^{2}\phi^{2}+\left(  \frac{3g^{2}}%
{2m^{2}}\phi^{4}+3\frac{g^{2}}{m^{4}}S_{2,2}-\frac{g^{2}}{2m^{4}}\right)
\nonumber \\
&  +\left(  -\frac{7}{6}\frac{g^{4}}{m^{6}}\phi^{6}-\frac{51}{2}\frac{g^{4}%
}{m^{8}}S_{2,4}-36\frac{g^{4}}{m^{10}}S_{4,2}+2\frac{g^{4}}{m^{12}}\pi
^{6}+\frac{15g^{4}}{2m^{8}}\phi^{2}+27\frac{g^{4}}{m^{10}}\pi^{2}\right)  ,
\label{herm1}%
\end{align}
where we set the mass parameter $m$ explicitly and the operator $(\nabla \phi)^{2}$
has been dropped out because it is certainly zero in $0+1$ dimensions. Also,
the symbol $S_{i,j}$ represents a symmetric combination of $j$ factors of the
field variable $\phi$ and $i$ factors of its conjugate field $\pi$. To
understand well the source of the divergent diagrams resulted in the
Hamiltonian in Eq.(\ref{herm1}), let us regroup the different terms in $h$ as;%
\begin{align}
h  &  =h_{0}+h_{f}+h_{sr}+h_{r}+h_{nr},\nonumber \\
h_{0}  &  =\frac{\pi^{2}}{2}+\frac{1}{2}m^{2}\phi^{2},\nonumber \\
h_{f}  &  =\frac{3g^{2}}{2m^{2}}\phi^{4}-\frac{7}{6}\frac{g^{4}}{m^{6}}%
\phi^{6}-\frac{g^{2}}{2m^{4}},\\
h_{sr}  &  =\frac{15g^{4}}{2m^{8}}\phi^{2}+3\frac{g^{2}}{m^{4}}S_{2,2}%
-\frac{51}{2}\frac{g^{4}}{m^{8}}S_{2,4},\nonumber \\
h_{r}  &  =-36\frac{g^{4}}{m^{10}}S_{4,2}+27\frac{g^{4}}{m^{10}}\pi
^{2},\nonumber \\
h_{nr}  &  =2\frac{g^{4}}{m^{12}}\pi^{6},\nonumber
\end{align}
where $h_{0}$ is the free Hamiltonian. Since the superficial degree of
divergence of a theory depends crucially on the mass dimension of the coupling
\cite{Peskin}, we realize that, in $0+1$ space-time dimensions, the couplings
in $h_{f}$ have mass dimension greater than $2$ and thus any diagram generated
from the contractions of the fields in $h_{f}$ is finite. For $h_{sr}$, the
couplings dimensions are positive but less than or equal $2$ which means that
this term represents a super renorlaizable interaction Hamiltonian and thus
one might find finite as well as infinite Feynman diagrams generated by
$h_{sr}$. Similarly, $h_{r}$ is renormalizable (has a dimensionless coupling)
while $h_{nr}$ is non-Renormalizable ( has a coupling of negative mass
dimension) interaction Hamiltonians even in $0+1$ space-time dimensions
(quantum mechanics). In view of these analysis, our important realization that
the results in Ref. \cite{bendvs} show an equivalence between finite
(non-Hermitian) and infinite \ (Hermitian) theories leaded us to believe that
the superficial degree of divergence is representation dependent.

For more elaboration of the possible equivalence between two theories of
different superficial degrees of divergence, in this work, we introduce and
study a class of non-Hermitian field theories for which the corresponding
equivalent Hermitian Hamiltonians are obtained in a closed form. The Hermitian
class of Hamiltonians has superficial degree of divergences which are different
from those of the corresponding non-Hermitian class of Hamiltonians. By this
work, we want to spread the message that one can play on the dimension (in
terms of mass) of the coupling in a theory and gets an equivalent theory with
a coupling of different mass dimension.

To start, let us consider the Hamiltonian model of the form%
\begin{align}
H  &  =\int dx^{3}\left(
\begin{array}
[c]{c}%
\frac{\pi^{2}\left(  x\right)  }{2}+\frac{1}{2}\left(  \nabla \phi \left(
x\right)  \right)  ^{2}+\frac{m^{2}}{2}\phi^{2}\left(  x\right) \\
+\frac{\lambda}{4!}\phi^{4}\left(  x\right)  +\frac{i\xi}{\sqrt{6!}}\left \{
\phi^{3}\left(  x\right)  ,\pi \left(  x\right)  \right \}
\end{array}
\right)  ,\nonumber \\
&  =H_{0}+\epsilon H_{I},\nonumber \\
H_{0}  &  =\int dx^{3}\left(  \frac{\pi^{2}\left(  x\right)  }{2}+\frac{1}%
{2}\left(  \nabla \phi \left(  x\right)  \right)  ^{2}+\frac{m^{2}}{2}\phi
^{2}\left(  x\right)  \right)  ,\label{model6f}\\
H_{I}  &  =\int dx^{3}\left(  \frac{\lambda}{4!}\phi^{4}\left(  x\right)
+\frac{i\xi}{2\sqrt{6!}}\left \{  \phi^{3}\left(  x\right)  ,\pi \left(
x\right)  \right \}  \right)  ,\nonumber
\end{align}
where $\phi$ is a one component scalar field, $\pi$ is its conjugate momenta
and $\left \{  A,B\right \}  $ is the anticommutator of two operators $A$ and
$B$. Also, $\epsilon$ is a parameter which can be set to one at the end. Note
that $\phi$ and $\pi$ are satisfying the commutation relations $\left[
\phi \left(  x\right)  ,\pi \left(  y\right)  \right]  =i\delta^{3}(x-y)$ and
$\left[  \phi \left(  x\right)  ,\phi \left(  y\right)  \right]  =\left[
\pi \left(  x\right)  ,\pi \left(  y\right)  \right]  =0$.

The Hamiltonian model in Eq.(\ref{model6f}) is non-Hermitian but according to
Mostafazadeh, if there exists a positive definite metric operator $\eta$ such
that $\eta H\eta^{-1}=H^{\dagger}$, then the spectrum of $H$ is real
\cite{spect,spect1}. Note also that, if $\eta$ exists then there exists an
equivalent Hermitian Hamiltonian operator $h$ such that;%
\begin{equation}
h=\rho H\rho^{-1},
\end{equation}
where $\rho=\sqrt{\eta}$. In using the relation $\eta H\eta^{-1}=H^{\dagger}$
and employing the form $\eta=\exp \left(  -Q\right)  $ we get the result;
\begin{align*}
H^{\dagger}  &  =\int dy^{3}\left(  \exp \left(  -Q\left(  y\right)  \right)
H\left(  x\right)  \exp(Q\left(  y\right)  )\right) \\
&  =\left(
\begin{array}
[c]{c}%
H\left(  x\right)  +\int dy^{3}[-Q\left(  y\right)  ,H\left(  x\right)
]+\frac{1}{2!}\int dy^{3}\int dz^{3}[-Q\left(  y\right)  ,[-Q\left(  z\right)
,H\left(  x\right)  ]]\\
+\frac{1}{3!}\int dy^{3}\int dz^{3}\int d\nu^{3}[-Q\left(  y\right)
,[-Q\left(  z\right)  ,[-Q\left(  \nu \right)  ,H\left(  x\right)  ]]]+....
\end{array}
\right)  ,
\end{align*}
where
\[
Q\left(  y\right)  =Q_{0}\left(  y\right)  +\epsilon Q_{1}\left(  y\right)
+\epsilon^{2}Q_{2}\left(  y\right)  ++\epsilon^{3}Q_{3}\left(  y\right)  +..
\]
Explicitly we have;
\begin{align*}
\exp(-Q)H\exp(Q)  &  =H_{0}+\epsilon H_{I}+[-Q,H_{0}]+[-Q,\epsilon
H_{I}]+\frac{1}{2!}[-Q,[-Q,H_{0}]]+\\
\frac{1}{2!}  &  [-Q,[-Q,\epsilon H_{I}]]+\frac{1}{3!}[-Q,[-Q,[-Q,H_{0}%
]]+\frac{1}{3!}[-Q,[-Q,[-Q,\epsilon H_{I}]]...\\
&  =H_{0}+\epsilon H_{I}^{\dagger},
\end{align*}

and thus
\begin{align}
0  &  =[-Q_{0},H_{0}]\text{ }\Rightarrow \text{ }Q_{0}=0\  \text{ is a good
choice.}\nonumber \\
H_{n}  &  =\frac{1}{2}[Q_{1},H_{0}]\nonumber \\
0  &  =[-Q_{2},H_{0}]+[-Q_{1},H_{I}]+\frac{1}{2!}[Q_{1},[Q_{1},H_{0}%
]]\nonumber \\
0  &  =[-Q_{3},H_{0}]+[-Q_{2},H_{I}]+\frac{1}{2!}[Q_{2},[Q_{1},H_{0}%
]]\nonumber \\
&  +\frac{1}{2!}[Q_{1},[Q_{2},H_{0}]]+\frac{1}{3!}[-Q_{1},[-Q_{1}%
,[-Q_{1},H_{0}]]]\\
&  +\frac{1}{3!}[-Q_{1},[-Q_{1},H_{I}]],\nonumber
\end{align}
where $H_{n}$ is the non-Hermitian term in the interaction Hamiltonian $H_{I}%
$.To get a Hermitian representation for the model $H=H_{0}+\epsilon H_{I}$,
one search for transformations which are able to kill the non-Hermitian
interaction term $H_{I}$. In fact, the assumption that $Q(\phi)$ is a real
functional of $\phi$ only will do the job because then the transformation of
$H_{0}$ with a suitable choice of $Q(\phi)$ will result in another functional
of $\phi$ times $\pi$. Considering this, one can expect a great simplification
to the above set of coupled operator equations. To show this, consider the
transformation of $H_{I}$;%

\[
\exp \left(  -Q\right)  H_{I}\exp \left(  Q\right)  ,
\]
since $H_{I}$ is linear in $\pi$ then the commutators $[Q_{n},H_{I}]$ are all
functionals in $\phi$ only. Accordingly, the above set will take the form;
\begin{align}
\text{ }Q_{0}  &  =0,\nonumber \\
H_{n}  &  =\frac{1}{2}[Q_{1},H_{0}],\\
Q_{2}  &  =Q_{3}=Q_{4}.....=0.\nonumber
\end{align}
Now, $Q_{1}$ is an operator when commuted with $H_{0}\left(  x\right)  $ gives
$2\int dx^{3}\frac{i\xi}{\sqrt{6!}}\left \{  \phi^{3}\left(  x\right)
,\pi \left(  x\right)  \right \}  $, then one can expect $Q_{1}$ to take the
form;
\begin{equation}
Q_{1}=\int dy^{3}\frac{\xi \phi^{4}\left(  y\right)  }{2\times \sqrt{6!}}.
\end{equation}
Thus,%
\begin{equation}
\eta=\exp \left(  -\int dy^{3}\frac{\xi \phi^{4}\left(  y\right)  }{2\times
\sqrt{6!}}\right)
\end{equation}
To obtain the equivalent Hermitian Hamiltonian consider the operator $\rho$
such that;
\begin{equation}
\rho=\exp \left(  -\int dy^{3}\frac{\xi \phi^{4}\left(  y\right)  }{4\times
\sqrt{6!}}\right)  =\exp \left(  -\int dy^{3}\omega \left(  y\right)  \right)  ,
\end{equation}

where we put $\omega(y)=\frac{\xi \phi^{4}\left(  y\right)  }{4\times \sqrt{6!}%
}$.

Then we have the form;
\[
\rho H\left(  x\right)  \rho^{-1}=H\left(  x\right)  +\int dy^{3}%
[-\omega \left(  y\right)  ,H\left(  y\right)  ]+\frac{1}{2}\int dz^{3}\int
dy^{3}[-\omega \left(  z\right)  ,[-\omega \left(  y\right)  ,H\left(  x\right)
]]+......
\]
In fact, only terms which have $\pi$ will be effective in calculating the
commutators and thus we have the result;
\begin{align}
\lbrack-\omega \left(  y\right)  ,\frac{\pi^{2}\left(  x\right)  }{2}]  &
=[-\omega \left(  y\right)  ,\pi \left(  x\right)  ]\frac{\pi \left(  x\right)
}{2}+\frac{\pi \left(  x\right)  }{2}[-\omega \left(  y\right)  ,\pi \left(
x\right)  ]\nonumber \\
&  =-i\frac{\partial \omega \left(  y\right)  }{\partial \phi \left(  y\right)
}\frac{\pi \left(  x\right)  }{2}\delta^{3}(x-y)-i\frac{\pi \left(  x\right)
}{2}\frac{\partial \omega \left(  y\right)  }{\partial \phi \left(  y\right)
}\delta^{3}(x-y)\\
&  =-\frac{1}{2}i\frac{\xi}{\sqrt{6!}}\left \{  \phi^{3}\left(  y\right)
,\pi \left(  x\right)  \right \}  \delta^{3}(x-y),\nonumber
\end{align}
and
\begin{align}
\left[  -\omega \left(  z\right)  ,[-\omega \left(  y\right)  ,\frac{\pi
^{2}\left(  x\right)  }{2}]\right]   &  =[-\omega \left(  z\right)  ,-\frac
{1}{2}i\frac{\xi}{\sqrt{6!}}\left \{  \phi^{3}\left(  y\right)  ,\pi \left(
x\right)  \right \}  \delta^{3}(x-y)]\nonumber \\
&  =2\frac{i}{2}\left(  \frac{\xi}{\sqrt{6!}}\phi^{3}\left(  y\right)
\right)  i\frac{\partial \omega \left(  z\right)  }{\partial \phi \left(
z\right)  }\delta^{3}(x-y)\delta^{3}(x-z)\nonumber \\
&  =-\left(  \frac{\xi}{\sqrt{6!}}\phi^{3}\left(  y\right)  \right)
^{2}\delta^{3}(x-y)\delta^{3}(x-z).
\end{align}
Then
\[
\rho \left(  \int d^{3}x\frac{\pi^{2}(x)}{2}\right)  \rho^{-1}=\int
d^{3}x\left(  \frac{\pi^{2}(x)}{2}-\frac{1}{2}i\left \{  \frac{\xi}{\sqrt{6!}%
}\phi^{3}\left(  x\right)  ,\pi \left(  x\right)  \right \}  -\frac{1}{2}\left(
\frac{\xi}{\sqrt{6!}}\phi^{3}\left(  x\right)  \right)  ^{2}\right)  .
\]
Similarly, we get%
\[
\rho H_{I}\rho^{-1}=H_{I}+\int dy^{3}[-\omega \left(  y\right)  ,H_{n}\left(
x\right)  ].
\]
Now we have;%
\begin{align}
\lbrack-\omega \left(  y\right)  ,H_{n}\left(  x\right)  ]  &  =[-\omega \left(
y\right)  ,i\frac{\xi}{2\sqrt{6!}}\left \{  \phi^{3}\left(  x\right)
,\pi \left(  x\right)  \right \}  ]\nonumber \\
&  =2\left(  i\frac{\xi}{2\sqrt{6!}}\phi^{3}\left(  x\right)  \right)
[-\omega \left(  y\right)  ,\pi \left(  x\right)  ]\nonumber \\
&  =2\left(  i\frac{\xi}{2\sqrt{6!}}\phi^{3}\left(  x\right)  \right)  \left(
-i\frac{\partial \omega \left(  y\right)  }{\partial \phi \left(  y\right)
}\right)  \delta^{3}(x-y)\\
&  =\left(  \frac{\xi}{\sqrt{6!}}\phi^{3}\left(  x\right)  \right)  ^{2}%
\delta^{3}(x-y).\nonumber
\end{align}
Collecting the different terms one gets the Hermitian Hamiltonian of the form;%
\begin{align}
h  &  =\int dx^{3}\left(
\begin{array}
[c]{c}%
\frac{1}{2}\left(  \nabla \phi \left(  x\right)  \right)  ^{2}+\frac{\pi
^{2}\left(  x\right)  }{2}+\frac{1}{2}m^{2}\allowbreak \phi^{2}\left(
x\right)  +\frac{g}{4!}\phi^{4}\left(  x\right) \\
-\frac{1}{2}\left(  \frac{\xi}{\sqrt{6!}}\phi^{3}\left(  x\right)  \right)
^{2}+\left(  \frac{\xi}{\sqrt{6!}}\phi^{3}\left(  x\right)  \right)  ^{2}%
\end{array}
\right) \nonumber \\
&  =\int dx^{3}\left(  \frac{1}{2}\left(  \nabla \phi \left(  x\right)  \right)
^{2}+\frac{\pi^{2}}{2}+\frac{1}{2}m^{2}\allowbreak \phi^{2}\left(  x\right)
+\frac{g}{4!}\phi^{4}\left(  x\right)  +\frac{1}{2}\left(  \frac{\xi}%
{\sqrt{6!}}\phi^{3}\left(  x\right)  \right)  ^{2}\right) \label{herm}\\
&  =\int dx^{3}\left(  \frac{\pi^{2}\left(  x\right)  }{2}+\frac{1}{2}\left(
\nabla \phi \left(  x\right)  \right)  ^{2}+\frac{1}{2}m^{2}\allowbreak \phi
^{2}\left(  x\right)  +\frac{g}{4!}\phi^{4}\left(  x\right)  +\frac{1}%
{2}\left(  \frac{\xi}{6!}\right)  ^{2}\phi^{6}\right)  .\nonumber
\end{align}
This form of the equivalent Hermitian Hamiltonian assures the real spectrum of
the non-Hermitian Hamiltonian model in Eq.(\ref{model6f}). The most
interesting realization is that the mass dimension of the couplings of the
interaction Hamiltonians in both the equivalent models in Eqs.(\ref{model6f}%
,\ref{herm}) are different. In Eq.(\ref{model6f}), the coupling of the
non-Hermitian interaction Hamiltonian $\xi$ has a mass dimension of the form
$M^{-1}$ in $3+1$ dimensions, where $M$ is a mass unit. On the other hand, the
Hermitian Hamiltonian in Eq.(\ref{herm}), the $\phi^{6}$ term has \ a coupling
$\varpropto \xi^{2}$ which has a mass dimensions of $-2$ in $3+1$ dimensions.
This is a very interesting result because the dimensionality of the coupling
constant determines whether the theory is super-renormalizable ( coupling of
positive mass dimension), renormalizable (dimensionless coupling) and
non-renormalizable (coupling of negative mass dimension). However, our result
showed that the coupling dimensions of some theory is representation
dependent. Relying on this closed form result and the perturbative
calculations in Ref .\cite{bendvs}, one can legally aim to find a
representation for a non-renormalizable theory in which the theory is
renormalizable. This kind of future research is very interesting toward the
unification of the four forces in our universe as gravity has a coupling of
negative $2$ mass dimension. Moreover, the $\phi^{6}$ theory which has a
coupling of negative $2$ mass dimension seems to be the only candidate to play
a role in early universe studies to account for a strong first order phase
transition needed for the matter-antimatter asymmetry in our universe. In view
of our work and the results in Ref.\cite{bendvs}, we think that
non-renormalizability of a theory is not a conceptual but a technical problem
that one can (in principle) get rid of it.

The non-Hermitian form studied above has another advantage. In fact, the
$\phi^{6}$ theory is a very interesting quantum field model for critical
phenomena studies and also for the search for models that have bound states.
However, the quantum field calculations for this model is not easy as each
vertex \ has an emergence of six lines and thus has a lengthy type of
calculations. On the other hand, the equivalent non-Hermitian form in
Eq.(\ref{model6f}) is a $\phi^{4}-like$ theory with only four lines emerging
at each vertex and thus the calculations in this representation goes more
easily than in the Hermitian representation.

To make the above studies more general, we consider the scalar quantum field
Hamiltonian of the form;%

\begin{equation}
H=\int dx^{3}\left(  \frac{\pi^{2}\left(  x\right)  }{2}+\frac{1}{2}\left(
\nabla \phi \left(  x\right)  \right)  ^{2}+m\frac{\phi^{2}\left(  x\right)
}{2}+i\left \{  G[\phi \left(  x\right)  ],\pi \left(  x\right)  \right \}
\right)  ,
\end{equation}
where $G[\phi \left(  x\right)  ]$ is a functional in the scalar field $\phi$.
The form of $\eta_{+}$ may be expected to have the form;
\begin{align}
\eta_{+}  &  =\rho^{2},\nonumber \\
\rho &  =\int dy^{3}\exp \left(  -\omega \lbrack \phi \left(  y\right)  ]\right)
,
\end{align}
where $\omega \lbrack \phi \left(  x\right)  ]$ is a functional in the scalar
field $\phi$ to be obtained. Using Baker-Campbell-Hausdorff formula we
obtain,
\begin{equation}
\rho H\rho^{-1}=H+\int dy^{3}[-\omega \left(  y\right)  ,H]+\int dy^{3}\int
dz^{3}\frac{1}{2}[-\omega \left(  z\right)  ,[-\omega \left(  y\right)
,H]]+...... \label{pi2w}%
\end{equation}
Now%
\begin{align*}
\lbrack-\omega \left(  y\right)  ,\frac{\pi^{2}\left(  x\right)  }{2}]  &
=[-\omega \left(  y\right)  ,\pi \left(  x\right)  ]\frac{\pi \left(  x\right)
}{2}+\frac{\pi \left(  x\right)  }{2}[-\omega \left(  y\right)  ,\pi \left(
x\right)  ],\\
&  =\left(  -i\frac{\partial \omega \left(  y\right)  }{\partial \phi \left(
y\right)  }\frac{\pi \left(  x\right)  }{2}-i\frac{\pi \left(  x\right)  }%
{2}\frac{\partial \omega \left(  y\right)  }{\partial \phi \left(  y\right)
}\right)  \delta^{3}\left(  x-y\right)  ,
\end{align*}
and
\begin{align}
\left[  -\omega \left(  z\right)  ,\left[  -\omega \left(  y\right)  ,\frac
{\pi^{2}\left(  x\right)  }{2}\right]  \right]   &  =\left[  -\omega \left(
z\right)  ,-i\left(  \frac{\partial \omega \left(  y\right)  }{\partial
\phi \left(  y\right)  }\frac{\pi \left(  x\right)  }{2}+\frac{\pi \left(
x\right)  }{2}\frac{\partial \omega \left(  y\right)  }{\partial \phi \left(
y\right)  }\right)  \right]  \delta^{3}\left(  x-y\right)  ,\nonumber \\
&  =-\frac{1}{2}\left(  \frac{\partial \omega \left(  z\right)  }{\partial
\phi \left(  z\right)  }\frac{\partial \omega \left(  y\right)  }{\partial
\phi \left(  y\right)  }+\frac{\partial \omega \left(  z\right)  }{\partial
\phi \left(  z\right)  }\frac{\partial \omega \left(  y\right)  }{\partial
\phi \left(  y\right)  }\right)  \delta^{3}\left(  x-z\right)  \left(
x-z\right)  .
\end{align}
Therefore,%
\begin{equation}
\left[  -\omega \left(  Q\right)  ,\left[  -\omega \left(  z\right)  ,\left[
-\omega \left(  y\right)  ,\frac{\pi^{2}\left(  x\right)  }{2}\right]  \right]
\right]  =0,
\end{equation}
and all the subsequent terms generated from the use of
Baker-Campbell-Hausdorff formula used for the commutation of $ \omega \left(  Q\right)$ with $\frac{\pi^{2}\left(  x\right)  }%
{2}$ are zero too. In other words, the closed form transformation of $H_{0}$
yields
\begin{align}
\rho H_{0}\rho^{-1}  &  =H_{0}+\int d^{3}x\int d^{3}y\left(  -i\frac
{\partial \omega \left(  y\right)  }{\partial \phi \left(  y\right)  }\frac
{\pi \left(  x\right)  }{2}-i\frac{\pi \left(  x\right)  }{2}\frac
{\partial \omega \left(  y\right)  }{\partial \phi \left(  y\right)  }\right)
\delta^{3}\left(  x-y\right) \nonumber \\
&  +\frac{1}{4}\int d^{3}x\int d^{3}y\int d^{3}z\left(  -\frac{\partial
\omega \left(  z\right)  }{\partial \phi \left(  z\right)  }\frac{\partial
\omega \left(  y\right)  }{\partial \phi \left(  y\right)  }-\frac{\partial
\omega \left(  z\right)  }{\partial \phi \left(  z\right)  }\frac{\partial
\omega \left(  y\right)  }{\partial \phi \left(  y\right)  }\right)  \delta
^{3}\left(  x-z\right)  \delta^{3}\left(  x-z\right) \\
&  =H_{0}-\frac{1}{2}i\int d^{3}x\left \{  \frac{\partial \omega \left(
x\right)  }{\partial \phi \left(  x\right)  },\pi \left(  x\right)  \right \}
-\frac{1}{2}\int d^{3}x\left(  \frac{\partial \omega \left(  x\right)
}{\partial \phi \left(  x\right)  }\right)  ^{2}.\nonumber
\end{align}
Now let us consider the transformation of $H_{I}$ and noting that;
\begin{align}
\left[  -\omega \left(  y\right)  ,\left \{  iG[\phi \left(  x\right)
],\pi \left(  x\right)  \right \}  \right]   &  =-i\left(  \left[  \omega \left(
y\right)  ,G[\phi \left(  x\right)  ]\pi \left(  x\right)  \right]  +\left[
\omega \left(  y\right)  ,\pi \left(  x\right)  G[\phi \left(  x\right)
]\right]  \right)  ,\nonumber \\
&  =2G[\phi \left(  x\right)  ]\left(  \frac{\partial \omega \left(  y\right)
}{\partial \phi \left(  y\right)  }\right)  \delta^{3}\left(  x-y\right)  ,
\end{align}
Then
\begin{align}
\rho H_{I}\rho^{-1}  &  =H_{I}+2\int d^{3}x\int d^{3}yG[\phi \left(  x\right)
]\left(  \frac{\partial \omega \left(  y\right)  }{\partial \phi \left(  y\right)
}\right)  \delta^{3}\left(  x-y\right)  ,\nonumber \\
&  =H_{I}+2\int d^{3}xG[\phi \left(  x\right)  ]\left(  \frac{\partial
\omega \left(  x\right)  }{\partial \phi \left(  x\right)  }\right)  ,
\end{align}

Accordingly,%
\begin{align}
\rho H\rho^{-1}  &  =h=H-\frac{1}{2}i\int d^{3}x\left \{  \frac{\partial
\omega \left(  x\right)  }{\partial \phi \left(  x\right)  },\pi \left(  x\right)
\right \}  -\frac{1}{2}\int d^{3}x\left(  \frac{\partial \omega \left(  x\right)
}{\partial \phi \left(  x\right)  }\right)  ^{2}\nonumber \\
&  +2\int d^{3}xG[\phi \left(  x\right)  ]\left(  \frac{\partial \omega \left(
x\right)  }{\partial \phi \left(  x\right)  }\right) \nonumber \\
&  =\int dx^{3}\left(  \frac{\pi^{2}\left(  x\right)  }{2}+\frac{1}{2}\left(
\nabla \phi \left(  x\right)  \right)  ^{2}+m\frac{\phi^{2}\left(  x\right)
}{2}+i\left \{  G[\phi \left(  x\right)  ],\pi \left(  x\right)  \right \}
\right)  ,\nonumber \\
&  -\frac{1}{2}i\int d^{3}x\left \{  \frac{\partial \omega \left(  x\right)
}{\partial \phi \left(  x\right)  },\pi \left(  x\right)  \right \}  -\frac{1}%
{2}\int d^{3}x\left(  \frac{\partial \omega \left(  x\right)  }{\partial
\phi \left(  x\right)  }\right)  ^{2}\\
&  +2\int d^{3}xG[\phi \left(  x\right)  ]\left(  \frac{\partial \omega \left(
x\right)  }{\partial \phi \left(  x\right)  }\right)  .\nonumber
\end{align}
In fact, if $G[\phi \left(  x\right)  ]=\frac{1}{2}\frac{\partial \omega \left(
x\right)  }{\partial \phi \left(  x\right)  }$, then $h$ is Hermitian and the
operator $\eta=\rho^{2}$ serves as a positive definite metric operator. To
check this, one uses the relation
\begin{equation}
H^{\dagger}=\eta H\eta^{-1}=\rho \rho H\rho^{-1}\rho^{-1}=\rho h\rho^{-1}.
\end{equation}
Or%
\begin{align}
H^{\dagger}  &  =\rho h\rho^{-1}=\int dx^{3}\rho \frac{\pi^{2}\left(  x\right)
}{2}\rho^{-1}+\int dx^{3}\left(  \frac{1}{2}\left(  \nabla \phi \left(
x\right)  \right)  ^{2}+m\frac{\phi^{2}\left(  x\right)  }{2}\right)
\nonumber \\
&  -\frac{1}{2}\int d^{3}x\left(  \frac{\partial \omega \left(  x\right)
}{\partial \phi \left(  x\right)  }\right)  ^{2}+2\int d^{3}xG[\phi \left(
x\right)  ]\left(  \frac{\partial \omega \left(  x\right)  }{\partial \phi \left(
x\right)  }\right)  ,\nonumber \\
&  =\int dx^{3}\left(  \frac{\pi^{2}\left(  x\right)  }{2}+\frac{1}{2}\left(
\nabla \phi \left(  x\right)  \right)  ^{2}+m\frac{\phi^{2}\left(  x\right)
}{2}\right) \nonumber \\
&  -\frac{1}{2}i\int d^{3}x\left \{  \frac{\partial \omega \left(  x\right)
}{\partial \phi \left(  x\right)  },\pi \left(  x\right)  \right \}  -\frac{1}%
{2}\int d^{3}x\left(  \frac{\partial \omega \left(  x\right)  }{\partial
\phi \left(  x\right)  }\right)  ^{2}\\
&  -\frac{1}{2}\int d^{3}x\left(  \frac{\partial \omega \left(  x\right)
}{\partial \phi \left(  x\right)  }\right)  ^{2}+2\int d^{3}xG[\phi \left(
x\right)  ]\left(  \frac{\partial \omega \left(  x\right)  }{\partial \phi \left(
x\right)  }\right) \nonumber \\
&  =\int dx^{3}\left(  \frac{1}{2}\left(  \nabla \phi \left(  x\right)  \right)
^{2}+\frac{\pi^{2}\left(  x\right)  }{2}+m\frac{\phi^{2}\left(  x\right)  }%
{2}\right)  -i\int d^{3}x\left \{  G[\phi \left(  x\right)  ],\pi \left(
x\right)  \right \}  .\nonumber
\end{align}
Accordingly, the form $\eta=\int dy^{3}\exp \left(  -\omega \left(  y\right)
\right)  $ passes all the tests as a positive definite metric operator for the
pseudo Hermitian Hamiltonian of the scalar field theory of the form;
\[
H=\int dx^{3}\left(  \frac{\pi^{2}\left(  x\right)  }{2}+\frac{1}{2}\left(
\nabla \phi \left(  x\right)  \right)  ^{2}+m\frac{\phi^{2}\left(  x\right)
}{2}+i\left \{  G[\phi \left(  x\right)  ],\pi \left(  x\right)  \right \}
\right)  .
\]

Let us now consider some specific choices for the functional $G[\phi(x)]$:

\textbf{Case 1: $\phi^{4}$ equivalent}

In this case, the functionals $w[\phi \left(  x\right)  ]$ and $G[\phi \left(
x\right)  ]$ take the forms;
\begin{align*}
w[\phi \left(  x\right)  ]  &  =\frac{g\phi^{3}\left(  x\right)  }{3},\\
G[\phi \left(  x\right)  ]  &  =\frac{1}{2}\frac{\partial \omega \left(
x\right)  }{\partial \phi \left(  x\right)  }=\frac{1}{2}g\phi^{2}\left(
x\right)  .
\end{align*}
then
\begin{align}
h  &  =\int dx^{3}\left(  \frac{\pi^{2}\left(  x\right)  }{2}+\frac{1}%
{2}\left(  \nabla \phi \left(  x\right)  \right)  ^{2}+m\frac{\phi^{2}\left(
x\right)  }{2}+\frac{1}{2}ig\left \{  [\phi^{2}\left(  x\right)  ],\pi \left(
x\right)  \right \}  \right)  ,\nonumber \\
&  -\frac{1}{2}ig\int d^{3}x\left \{  \frac{1}{2}\phi^{2}\left(  x\right)
,\pi \left(  x\right)  \right \}  -\frac{1}{2}g^{2}\int d^{3}x\left(  \phi
^{2}\left(  x\right)  \right)  ^{2}+g^{2}\int d^{3}x\phi^{4}\left(  x\right)
,\nonumber \\
&  =\int dx^{3}\left(  \frac{\pi^{2}\left(  x\right)  }{2}+\frac{1}{2}\left(
\nabla \phi \left(  x\right)  \right)  ^{2}+m\frac{\phi^{2}\left(  x\right)
}{2}+\frac{1}{2}g^{2}\phi^{4}\left(  x\right)  \right)  .
\end{align}

\textbf{Case 2: $\phi^{6}$ equivalent}

In this case; we choose the functional $w[\phi \left(  x\right)  ]$ as
\begin{align*}
w[\phi \left(  x\right)  ]  &  =\frac{\lambda \phi^{4}\left(  x\right)  }{4},\\
G[\phi \left(  x\right)  ]  &  =\frac{1}{2}\frac{\partial \omega \left(
x\right)  }{\partial \phi \left(  x\right)  }=\frac{1}{2}\lambda \phi^{3}\left(
x\right)  ,
\end{align*}
and thus
\begin{align}
h  &  =\int dx^{3}\left(  \frac{\pi^{2}\left(  x\right)  }{2}+\frac{1}%
{2}\left(  \nabla \phi \left(  x\right)  \right)  ^{2}+m\frac{\phi^{2}\left(
x\right)  }{2}\right) \nonumber \\
&  -\frac{1}{2}\int d^{3}x\left(  \lambda \phi^{3}\left(  x\right)  \right)
^{2}+4\int d^{3}x\left(  \frac{\lambda \phi^{3}\left(  x\right)  }{2}\right)
^{2}\\
&  =\int dx^{3}\left(  \frac{1}{2}\left(  \nabla \phi \left(  x\right)  \right)
^{2}+\frac{\left(  \pi \left(  x\right)  \right)  ^{2}}{2}+m\frac{\phi
^{2}\left(  x\right)  }{2}+\frac{1}{2}\lambda^{2}\phi^{6}\left(  x\right)
\right)  .
\end{align}

\textbf{Case 3: $\phi^{8}$ equivalent}

\begin{figure}[htbp]
	\centering
		\includegraphics{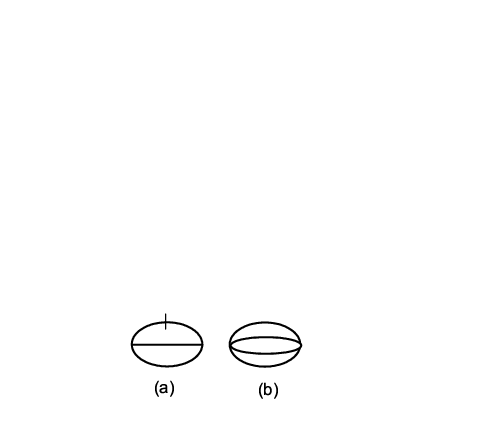}
	\caption{Two Feynman diagrams with the same number of vertices in the Non-Hermitian theory $i\left\{\phi^2,\pi\right\}$ (diagram $(a)$) and the  corresponding equivalent Hermitian $\phi^4$ theory (diagram $(b)$). The tick mark in one of the propagator of  diagram $(a)$ refers to the contraction of two $\pi$ fields.}
	\label{feyn}
\end{figure}

In this case we choose the functional $w[\phi \left(  x\right)  ]$ as
\begin{align*}
w[\phi \left(  x\right)  ] &  =\frac{\lambda_{8}\phi^{4}\left(  x\right)  }%
{4}+\frac{\lambda_{6}\phi^{3}\left(  x\right)  }{3}+\frac{\lambda_{4}\phi
^{2}\left(  x\right)  }{2},\\
G[\phi \left(  x\right)  ] &  =\frac{1}{2}\frac{\partial \omega \left(  x\right)
}{\partial \phi \left(  x\right)  }=\frac{1}{2}\left(  \lambda_{8}\phi
^{4}\left(  x\right)  +\lambda_{6}\phi^{3}\left(  x\right)  +\lambda_{4}%
\phi^{2}\left(  x\right)  \right)
\end{align*}
and thus
\begin{align}
h &  =\int dx^{3}\left(  \frac{\pi^{2}\left(  x\right)  }{2}+\frac{1}%
{2}\left(  \nabla \phi \left(  x\right)  \right)  ^{2}+m\frac{\phi^{2}\left(
x\right)  }{2}\right)  \nonumber \\
&  +\frac{1}{2}\int d^{3}x\left(  \lambda_{8}^{2}\phi^{8}+2\lambda_{6}%
\lambda_{8}\phi^{7}+\left(  \lambda_{6}^{2}+2\lambda_{4}\lambda_{8}\right)
\allowbreak \phi^{6}+2\lambda_{4}\lambda_{6}\phi^{5}\allowbreak+\lambda_{4}%
^{2}\phi^{4}\right)  .
\end{align}
What is interesting in this representation is that it lowers the number of
lines $n$ emerging from each vertex in the Hermitian theory to $\frac{n}{2}+1$
lines emerging from each vertex in the non-Hermitian representation. This
property lowers the number of available Feynman diagrams and thus simplifies
the perturbative calculations. To elucidate this point, consider the two
Feynman diagrams in Fig.\ref{feyn}. which they correspond to second order  (in the
couplings) of the equivalent Hermitian $\phi^4$ theory and non-Hermitian $i\phi^2 \pi$ theory. Both diagrams
have the same number of vertices  however the diagram that correspond to the
non-Hermitian theory diverge as $\Lambda^{2d-4}$ while the diagram that correspond to
the Hermitian $\phi^{4}$ theory diverge as $\Lambda^{3d-8}$. Moreover, in the
non-Hermitian theory, one deals with three propagators while for the
calculation of the  diagram in the Hermitian theory one deals with four
propagators. Accordingly, power counting shows that the UV behavior of the
non-Hermitian theory is better as well as the Feynman diagrams calculations
are simple provided that we compare diagrams of same number of vertices  in the
Hermitian and the equivalent non-Hermitian theories.

To conclude,  we asserted the non-existence of a no go theorem
for the possible physical acceptability of a non-renormalizable theory. To
show that, we revisited a previous work by Carl M. Bender, Jun-Hua Chen, and
Kimball A. Milton \cite{bendvs}. In this article, the authors found
divergences in the Feynman diagrams generated by the Hermitian Hamiltonian.
The importance of this result has been overlooked in the Benders
\textit{et.al} work as well as in the literature. Although the calculations in
Ref.\cite{bendvs} have been done in $0+1$ space-time dimensions, we used
dimensional analysis to show that the divergences found are due to
superrenormalizable, renormalizable and non-renormalizable interaction terms
in the Hermitian Hamiltonian. Since the Hermitian Hamiltonian is now
non-renormalizable while the equivalent non-Hermitian theory is finite, we
concluded the possibility of the dependence of the superficial degree of
divergence of a theory on its representation (e.g Hermitian and non-Hermitian
equivalent representations ). To elucidate the idea, we introduced a class of
non-Hermitian Hamiltonians and obtained the equivalent class of Hermitian
Hamiltonians in a closed form. For a non-Hermitian Hamiltonian in the class
that has a coupling of mass dimension like $M^{\frac{\delta}{2}}$, we realized
that the corresponding Hermitian Hamiltonian has a coupling of mass dimension
as $M^{\delta}$. Accordingly, we proved the possibility of having two
equivalent theories which have different superficial degree of divergences.
This is a very interesting result which may help in finding a solution to the
unification problem.

The Hermitian representation is of the form $\phi^{2n}$ scalar field theory while the
equivalent non-Hermitian representation is a $\phi^{n+1}-like$ theory which
turn the calculations more simpler in this representation than in the
Hermitian one. Besides, the metric operator is supposed to disappear from path
integral calculations which mean that physical amplitudes can be fully
obtained in the simpler non-Hermitian representation.


\begin{thebibliography}{99}                                                                                               %


\bibitem {Frieder}Frieder Kleefeld, J. Phys. A: Math. Gen. 39 L9--L15 (2006).

\bibitem {symanzic}Symanzik K, Springer Tracts Mod. Phys. 57 222--36 (1971).

\bibitem {symanzic1}Symanzik K, Commun. Math. Phys. 23 49--86 (1971).

\bibitem {symanzic2}Symanzik K, Nuovo Cimento 6 77--80 (1973).

\bibitem {bendr}Carl Bender and Stefan Boettcher, Phys.Rev.Lett.80:5243-5246 (1998).

\bibitem {aboebt}Abouzeid Shalaby and Suleiman S. Al-Thoyaib, Phys. Rev.D 82,
085013 (2010)

\bibitem {ghost}Abouzeid M. Shalaby, Phys.Rev.D80:025006 (2009).

\bibitem {ghost1}Carl M. Bender and Philip D. Mannheim,
Phys.Rev.Lett.100:110402 (2008).

\bibitem {ghost2}Carl M. Bender, Sebastian F. Brandt, Jun-Hua Chen and Qinghai
Wang, Phys.Rev. D71, 025014 (2005).

\bibitem {early}Edward W.Kolb and Michaels. Turner, The Early Universe,
ADDISON-WESLEY PUBLISHING COMPANY, (1990).

\bibitem {early2}Christophe Grojean, Geraldine Servant and James D. Wells,
Phys.Rev.D71:036001 (2005).

\bibitem {bendvs}Carl M. Bender, Jun-Hua Chen, and Kimball A. Milton,
J.Phys.A39:1657-1668 (2006).

\bibitem {Peskin}Michael E. Peskin and Daniel V.Schroeder, An Introduction To
Quantum Field Theory (Addison-Wesley Advanced Book Program) ( 1995).

\bibitem {spect}A. Mostafazadeh, J. Math. Phys., 43, 3944 (2002).

\bibitem {spect1}A. Mostafazadeh, J. Math. Phys. 43, 205 (2002).
\end{thebibliography}
\end{document}